\documentclass[aps,twocolumn,pra,superscriptaddress,showpacs,tightenlines]{revtex4-1}
\usepackage{amsmath}
\usepackage{graphicx}
\usepackage{color}
\usepackage{amsfonts}
\usepackage[colorlinks,citecolor=blue]{hyperref}
\begin{document}

\title{Enhancement of mechanical effects of single photons in modulated two-mode optomechanics}
\author{Jie-Qiao Liao}
\email{jieqiaoliao@gmail.com}
\affiliation{School of Natural
Sciences, University of California, Merced, California 95343, USA}
\affiliation{CEMS, RIKEN, Saitama 351-0198, Japan}
\author{C. K. Law}
 \affiliation{Department of Physics
and Institute of Theoretical Physics, The Chinese University of
Hong Kong, Shatin, Hong Kong Special Administrative Region, China}
\author{Le-Man Kuang}
\affiliation{Key Laboratory of Low-Dimensional Quantum Structures and Quantum Control of Ministry of Education,
and Department of Physics, Hunan Normal University, Changsha 410081, China}
\author{Franco Nori}
\affiliation{CEMS, RIKEN, Saitama 351-0198, Japan}
\affiliation{Department of Physics, The University of Michigan, Ann Arbor, Michigan 48109-1040, USA}
\date{\today}

\begin{abstract}
We propose an approach to enhance the mechanical effects of single
photons in a two-mode optomechanical system. This is achieved by
introducing a resonance-frequency modulation to the cavity fields.
When the modulation frequency and amplitude satisfy certain
conditions, the mechanical displacement induced by single photons
could be larger than the quantum zero-point fluctuation of the
oscillating resonator. This method can be used to create distinct
mechanical superposition states.
\end{abstract}


%42.50.Wk     Mechanical effects of light on material media, microstructures and particles
%42.50.Pq     Cavity quantum electrodynamics; micromasers
%42.50.Dv     Quantum state engineering and measurements

\maketitle

\section{Introduction}

One of the most interesting regimes in cavity quantum
optomechanics~\cite{Aspelmeyer2014,Chen2013} occurs when the
radiation pressure of a single photon is strong enough to push a
mechanical oscillator with a displacement greater than the
oscillator's zero-point fluctuations before escaping the cavity.
In such a regime, one can generate distinct superposition states
of a macroscopic object (the oscillator)~\cite{Bose1999,Marshall2003},
and the quantum nonlinearity
in the ultrastrong optomechanical coupling regime can lead to
various quantum effects, including photon
blockade~\cite{Rabl2011,Liao2013,Xu2013A,Kronwald2013},
non-Gaussian states of the mechanical
oscillator~\cite{Nunnenkamp2011,Xu2013B,Gaofeng2013,Tan2014,Garziano2015}, modified
spectra of photon scattering~\cite{Liao2012,Liao2013}, and
dressed-state-representation dissipations~\cite{Hu2015}.
Specifically, this regime is defined by the condition:
\begin{equation}
g_0 > \omega_M \gg \kappa_c,\label{regcondition1}
\end{equation}
where $g_0$ is the single-photon optomechanical coupling strength,
$\omega_M$ is the frequency of the
mechanical oscillator, and $\kappa_c$ is the cavity field damping
rate.

The requirement of $g_0> \omega_M$ in~(\ref{regcondition1}) is understood from a
generic single-mode time-independent optomechanical system with
the Hamiltonian ${\cal H} =\omega_{c} a^\dag a + \omega_{M} b^\dag b
-g_{0} a^\dag a(b+b^\dag)$, where $a$ $(b)$ and $a^\dag$ $(b^\dag)$
are the annihilation and creation operators of the cavity field
(mechanical oscillator). By solving the Schr\"odinger equation,
one finds that the oscillator, when driven by a single photon, can
evolve to a coherent state $|2g_{0}/\omega_{M}\rangle$ from the
ground state after a time $t= \pi/ \omega_{M}$. Therefore, $g_{0}>\omega_{M}$
provides a significant displacement away from the ground state.

However, the realization of condition~(\ref{regcondition1})
has been challenging for current experiments because $g_0$ is small
in most optomechanical systems, and $\omega_M$ has to be
sufficiently high to overcome thermal noise effects. In particular,
the quantum limit of phonon number $n_b$ obtained by resolved
sideband cooling is $n_b = \kappa_c^2/(16 \omega_M^2)$
\cite{Wilson-Rae2007,Marquardt2007,Li2008,Tian2009,Liu2015}. It
should be noted that condition~(\ref{regcondition1}) is based
on the single-mode time-independent Hamiltonian ${\cal H}$, and thus
an interesting question is whether it is possible to achieve
ultrastrong coupling effects in multimode and time-dependent
systems. Indeed, a similar question has been explored in
electromechanical systems recently~\cite{Liaomodu,Lu2013}, and it
is found that a suitable modulation can be used to effectively
enhance photonic nonlinearities in electromechanical systems. In
addition, an interferometric scheme has been proposed to detect
optomechanical coherent interaction at the single-photon
level~\cite{Tang2014}.

The methods of enhancing radiation pressure effects of a single
photon in optomechanical systems have been recently discussed by several
authors~\cite{Xuereb2012,Heikkila2014,Rimberg2014,Lu2015,Via2015}. In
this paper, we approach this problem by modulating the
photon-tunneling process in the ``membrane-in-the-middle" (MIM)
configuration. The MIM optomechanical system has been studied
experimentally~\cite{Thompson,Sankey,Karuza1,Karuza2,Regal1,Karuza3,Andrews2014,Lee,Regal2}
and
theoretically~\cite{Bhattacharya2008,Jayich2008,Miao2009,Cheung2011,Ludwig2012,Stannigel2012,Liao2014}.
Here we show that by changing the frequency of the cavities in a
time-dependent manner, the photon tunneling oscillation can be
modulated. We discover a set of conditions such that a single
photon can significantly displace the membrane, even though
$g_0/\omega_M$ is much less than $1$. Below we derive an effective
Hamiltonian of the system and present numerical evidence of the
ultrastrong coupling effect.

\section{The model}

\begin{figure}[tbp]
\center
\includegraphics[bb=130 657 310 732, width=0.48 \textwidth]{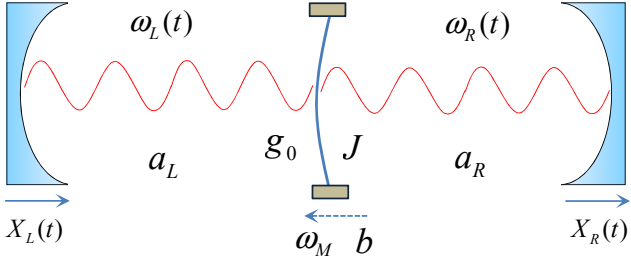}
\caption{(Color online) Schematic of the optomechanical system
with the ``membrane-in-the-middle" configuration. The cavity
fields couple with the mechanical oscillation via the radiation
pressure coupling $g_{0}$, and the optical fields in the left and right
cavities couple to each other via the photon-hopping
interaction $J$. To realize the cavity-frequency modulation in Eq.~(\ref{modulations}),
the two outside mirrors could be forced to oscillate with the same displacements
$X_{L}(t)=X_{R}(t)=(\Delta_{0}/\omega_{c})L\cos(2Jt)$ by using an external device such as a piezoelectric transducer.}\label{setup}
\end{figure}

We consider an MIM optomechanical system
(Fig.~\ref{setup}) in which a vibrational mode of the membrane is
coupled to the left and right cavity field modes via the radiation
pressure difference between the two cavities, and tunneling of
photons through the membrane is allowed. The Hamiltonian of the
system~\cite{Bhattacharya2008,Jayich2008,Miao2009,Cheung2011,Ludwig2012,Stannigel2012,Liao2014}
is modeled by ($\hbar=1$)
\begin{eqnarray}
H&=&\omega_{L} a_{L}^{\dagger}a_{L}+\omega_{R}a_{R}^{\dagger}a_{R}
+J(a_{L}^{\dagger }a_{R}+a_{R}^{\dagger }a_{L})\nonumber\\
&&+\omega_{M}b^{\dagger}b+g_{0}(a_{L}^{\dagger}a_{L}-a_{R}^{\dagger}a_{R})( b+b^{\dagger}),\label{Hamiltonian1}
\end{eqnarray}
where $a_{L(R)}$ and $b$ are the annihilation operators of the
left (right) cavity and the mechanical membrane, with the
respective resonance frequencies $\omega_{L(R)}$ and $\omega_{M}$.
The parameters $J$ and $g_{0}$ are the coupling strengths of the
photon hopping and the optomechanical interaction, respectively.

%%%%%%%%%%%%%%%%%%%%%%%%%%%%%%%%%%%%
%%%%%%%%%%%%%%%%%%%%%%%%%%%%%%%%%%%%

A key feature in our model is the introduction of a weak frequency
modulation to the left and the right cavities such that $\omega_L$
and $\omega_R$ in Eq.~(\ref{Hamiltonian1}) are given by
\begin{subequations}
\label{modulations}
\begin{align}
\omega_{L}(t)&=\omega_{c}+\Delta_{0}\cos(2Jt),\\
\omega_{R}(t)&=\omega_{c}-\Delta_{0}\cos(2Jt),
\end{align}
\end{subequations}
where the modulation amplitude $\Delta_0$ is assumed to be much
smaller than $J$. For simplicity, $\Delta_{0}$ and $J$ are taken
to be positive numbers. The sinusoidal time dependence of the cavity-field frequencies could be realized by modulating the lengths of the cavities. When the lengths of the left and right cavities change as $L-(\Delta_{0}/\omega_{c})L\cos(2Jt)$
and $L+(\Delta_{0}/\omega_{c})L\cos(2Jt)$, respectively, then the frequencies of the two cavities are modulated as in Eq.~(\ref{modulations}).
Here $L$ is the cavity length when the membrane and the outside mirrors are at rest. We assume that
the motion of two outside mirrors is prescribed classically, and the back-action on such mirrors is neglected.

To realize the modulation of the cavity lengths, one could use piezoelectric transducers, which convert the electrical signal to mechanical vibration~\cite{Svelto}. We could also mount the end mirrors on mechanical resonators, such as cantilevers, which oscillate periodically.
In addition, one could shake the outside mirrors by shining each
of them with two lasers of frequency difference $2J$, since the
beating between the two lasers yields a radiation pressure force
modulated at a frequency $2J$~\cite{Schmidt2015}.

The evolution of the system state $|\psi(t) \rangle$ is
governed by the Schr\"odinger equation $H(t)|\psi (t) \rangle = i
d| \psi (t) \rangle/dt $. To study how the frequency modulation
affects the quantum dynamics, it is useful to introduce the
transformation $|\psi (t) \rangle = T(t) |\tilde \psi (t)
\rangle$, where $T(t)$ is a unitary operator defined by $T(t)
\equiv V_1(t) V_2(t) V_3 (t)$ with
\begin{subequations}
\label{transfms}
\begin{align}
& V_{1}(t)=\exp[-i\omega_{c}t(
a_{L}^{\dagger}a_{L}+a_{R}^{\dagger}a_{R})],\label{defV1} \\
& V_{2}(t)=\exp[-iJt(a_{L}^{\dagger}a_{R}+a_{R}^{\dagger}a_{L})],\\
&V_{3}(t)=\exp\left\{-i\left[\frac{\Delta_{0}}{2}(a_{L}^{\dagger}a_{L}-a_{R}^{\dagger}a_{R})
+\omega_{M}b^{\dagger}b\right]t\right\}.
\end{align}
\end{subequations}
The evolution of $|\tilde \psi (t) \rangle$ is governed by the
transformed Hamiltonian $\tilde H(t)= T^{\dagger}(t)H T(t) -i
T^{\dagger }(t) {d T(t)}/ {dt}$, which has the form
\begin{eqnarray}
\label{Hamiltoscilterm}
&&\tilde H (t)\nonumber\\
&=&\frac{\Delta_{0}}{4}(a_{L}^{\dagger}a_{L}-a_{R}^{\dagger
}a_{R})(e^{4iJt}+e^{-4iJt})\notag \\
&&+\frac{\Delta_{0}}{4}[a_{R}^{\dagger}a_{L}(e^{i(4J-\Delta_{0})t}-e^{-i(4J+\Delta_{0})t})+\textrm{H.c.}]\notag\\
&&+\frac{g_{0}}{2}(a_{L}^{\dagger}a_{L}-a_{R}^{\dagger}a_{R})
(be^{-i(\omega_{M}-2J)t}+b^{\dagger}e^{i(\omega_{M}-2J)t})\notag\\
&&+\frac{g_{0}}{2}(a_{L}^{\dagger}a_{L}-a_{R}^{\dagger}a_{R})
(be^{-i(\omega_{M}+2J)t}+b^{\dagger}e^{i(\omega_{M}+2J)t})\notag \\
&&+\frac{g_{0}}{2}\left[a_{L}^{\dagger }a_{R}b^{\dagger
}(e^{i[\Delta _{0}+(\omega
_{M}-2J)]t}-e^{i[\Delta_{0}+(\omega_{M}+2J)]t})\right.\nonumber\\
&&\left.+a_{L}^{\dagger}a_{R}b(e^{i[\Delta_{0}-(\omega_{M}+2J)] t}-e^{i[\Delta_{0}-(\omega _{M}-2J)]t})+\textrm{H.c.}\right].\nonumber\\
\end{eqnarray}
The motivation for performing the above transformation is that the
parameters $\omega_M$ and $J$, which are typically
much greater than $g_0$ and $\Delta_0$, all appear in oscillating
phase factors. The form of $\tilde H (t)$ allows us to identify
fast oscillating terms and perform the rotating-wave approximation.
Specifically, under the conditions
\begin{eqnarray}
|\omega_{M}- 2J| \le \frac{g_0}{2},\hspace{0.5 cm}
J\gg\frac{5}{16}\Delta _{0},\hspace{0.5 cm} \Delta_{0}\gg
g_{0},\label{conditionRWA}
\end{eqnarray}
we may retain only the slowly oscillating terms
$\frac{g_{0}}{2}(a_{L}^{\dagger}a_{L}-a_{R}^{\dagger}a_{R})[be^{-i(\omega_{M}-2J)t}+b^{\dagger}e^{i(\omega_{M}-2J)t}]$
in Eq.~(\ref{Hamiltoscilterm}), and discard all other terms. This is justified within the
rotating-wave approximation, because in order to discard
those terms, their oscillation frequencies must be much larger than
their corresponding coupling coefficients under
conditions~(\ref{conditionRWA}). Hence Eq.~(\ref{Hamiltoscilterm}) is approximated by
\begin{eqnarray}
\tilde H (t)\approx
\frac{g_{0}}{2}(a_{L}^{\dagger}a_{L}-a_{R}^{\dagger}a_{R})
(be^{-i(\omega_{M}-2J)t}+ \textrm{H.c.}).\label{Hamiltappr1}
\end{eqnarray}
Such a Hamiltonian can be cast into a familiar form in a
rotating frame defined by $|\tilde \psi (t)\rangle = S(t) |\tilde \psi' (t)
\rangle$, with the unitary operator $S(t)
=\exp\{i[\omega_c (a_{L}^{\dagger}a_{L}+a_{R}^{\dagger}a_{R})
+(\omega_{M}-2J) b^{\dagger}b]t\} $. The transformed Hamiltonian $\tilde H'=S^{\dagger}(t) \tilde H(t)S(t) -i
S^{\dagger }(t) dS(t)/dt$ then reads
\begin{eqnarray}
\tilde H' & \approx &\omega_c
(a_{L}^{\dagger}a_{L}+a_{R}^{\dagger}a_{R})
+(\omega_{M}-2J) b^{\dagger}b  \nonumber \\
&& + \frac{g_{0}}{2}(a_{L}^{\dagger}a_{L}-a_{R}^{\dagger}a_{R})
(b+ b^\dag).\label{Hamiltappr2}
\end{eqnarray}
The right side is precisely the same form of the Hamiltonian of
the MIM optomechanical system without the
tunneling term, but with an effective mechanical frequency
$\omega_{M}-2J$, which is significantly reduced. Consequently, the
mechanical displacement induced by a single photon, which is now
proportional to $g_{0}/[2(\omega_{M}-2J)]$, can be significant.

As a remark, our model can be analyzed by working in the Schwinger representation. Define the angular momentum operators $S_{x}=(a_{L}^{\dag} a_{R}+a_{R}^{\dag} a_{L})/2$, $S_{y}=i(a_{R}^{\dag} a_{L}-a_{L}^{\dag} a_{R})/2$, and $S_{z}=(a_{L}^{\dag} a_{L}-a_{R}^{\dag} a_{R})/2$, then the Hamiltonian~(\ref{Hamiltonian1}), up to a constant-of-motion term $\omega_{c}(a_{L}^{\dag} a_{L}+a_{R}^{\dag} a_{R})$, can be expressed as
\begin{equation}
H=2\Delta _{0}\cos(2Jt)S_{z}+2JS_{x}+\omega_{M}b^{\dagger}b+2gS_{z}(b+b^{\dagger}),\label{schwingerrepH}
\end{equation}
which is the quantum Rabi model for $S=1/2$ (single-photon case)
and $\Delta_0 =0$. The $\Delta_0$ term corresponds to a driving term. It
is worth noting that the quantum Rabi model in the
ultrastrong-coupling regime, which is difficult to be realized in typical
systems, can be simulated by reducing the effective qubit
frequency~\cite{Ballester2012}.

\section{Single-photon mechanical displacement}

Let us consider an initial state of the system in which the
membrane is in the ground state and a photon is in the left cavity
mode,
\begin{equation}
\vert\psi(0)\rangle=\vert \tilde \psi'(0)\rangle=\vert
1\rangle_{{L}}\vert 0\rangle_{{R}}\vert 0\rangle_{M}\label{inistate}.
\end{equation}
Then, by using the approximate Hamiltonian~(\ref{Hamiltappr2}), we have
\begin{equation}
|\psi (t)\rangle \approx T(t) S(t) e^{-i\tilde H' t} \vert \tilde
\psi'(0)\rangle  \equiv |\psi_{\rm approx}(t) \rangle \label{defpsiapprox}
\end{equation}
after transforming back to the original frame. Explicitly, the
approximate state $|\psi_{\rm approx}(t) \rangle$ is given by
\begin{eqnarray}
|\psi_{\rm approx}(t) \rangle &=&e^{-i\theta(t)}e^{-i\omega_{c}t}[\cos(Jt)\vert 1\rangle_{{L}}\vert 0\rangle_{{R}}\nonumber\\
&&-i\sin(Jt)\vert
0\rangle_{{L}}\vert1\rangle_{{R}}]\vert\beta(t)\rangle_{M},\label{psit}
\end{eqnarray}
where $|\beta (t) \rangle_M$ is a coherent state of the membrane,
and $\theta(t)$ and $\beta (t)$ are given by
\begin{subequations}
\begin{align}
\theta(t)&=\left(\frac{\Delta_{0}}{2}-\frac{g_{0}^{2}}{4(\omega_{M}-2J)}\right)t+\frac{g_{0}^{2}\sin[(\omega _{M}-2J)t]}{4(\omega_{M}-2J)^{2}},\\
\beta(t)&=\frac{g_{0}}{2(\omega_{M}-2J)}(e^{-i\omega_{M}t}-e^{-2iJt}).\label{detaoft}
\end{align}
\end{subequations}
Hence the membrane can reach a maximum displacement
${g_{0}}/{(\omega_{M}-2J)}$ at the time $t_s=\pi/(\omega_{M}-2J)$,
assuming $\omega_M > 2J$.  A large displacement amplitude of the
membrane can be obtained if $2J$ is close to $\omega_M$. In
particular, at $2J=\omega_M$, the displacement amplitude in
Eq.~(\ref{detaoft}) takes the form:
$\beta(t)=-ig_{0}t\exp(-i\omega_{M}t)/2$, which grows linearly in
time until it becomes too large that
Hamiltonian~(\ref{Hamiltonian1}) breaks down. This is in contrast
to unmodulated systems (i.e., $\Delta_0=0$) in which the membrane
can gain less than one phonon for the same initial
condition~\cite{jc}.
%%%%%%%%%%%%%%%%%%%%%%%%%%%%%%%%%%%
\begin{figure}[tbp]
\center
\includegraphics[bb=5 11 347 424, width=0.48 \textwidth]{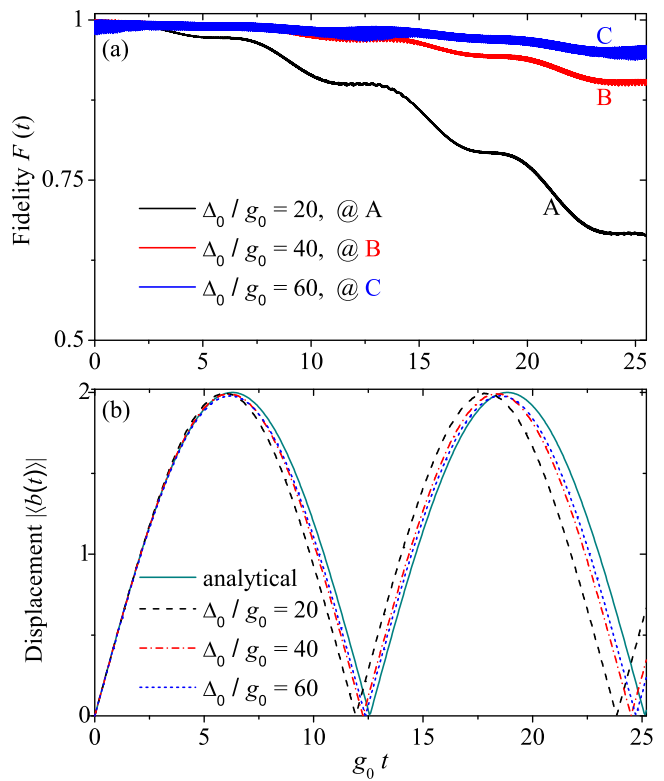}
\caption{(Color online) (a) The fidelity $F(t)$ versus the scaled
time $g_{0}t$ when the parameter $\Delta_{0}/g_{0}$ takes various
values. (b) Comparison of the mechanical displacements $|\langle
b(t)\rangle|$ in the numerical case (for $\Delta_{0}/g_{0}=20$,
$40$, and $60$) with that in the analytical case. Other parameters
are given by $\omega_{M}/g_{0}=201$ and
$J/g_{0}=100.25$.}\label{fidelity}
\end{figure}
%%%%%%%%%%%%%%%%%%%%%%%%%%%%%%%%%%%

To check the validity of our approximation, we performed a numerical
calculation of the state $|\psi(t)\rangle$ with the Hamiltonian $H$
in Eq.~(\ref{Hamiltonian1}) subjected to the frequency modulation
defined by Eq.~(\ref{modulations}) and the same initial condition as above. The numerically exact $|
\psi (t) \rangle$ is then compared with the $|\psi_{\rm approx}(t)
\rangle$ of Eq.~(\ref{psit}). Specifically, we examine the fidelity $F(t)
= \vert\langle \psi(t)\vert \psi_{\rm approx}(t)\rangle\vert^{2}$.

We remark that in solving numerically the Schr\"{o}dinger
equation using the Hamiltonian~(\ref{Hamiltonian1}) and the initial state
$\vert\psi (0)\rangle=\vert 1\rangle_{L}\vert
0\rangle_{R}\vert 0\rangle_{M}$, we express the system state
at later time as \begin{eqnarray} \vert\psi
(t)\rangle=\sum_{m=0}^{\infty}[A_{m}(t)\vert 1\rangle_{{L}}\vert
0\rangle_{{R}}+B_{m}(t)\vert 0\rangle_{{L}}\vert 1\rangle_{{R}}]\vert m\rangle_{M}.\label{psiprimet}
\end{eqnarray}
Then, by the Schr\"{o}dinger equation, the probability amplitudes
$A_{m}$ and $B_{m}$ are governed by the following set of coupled differential
equations:
\begin{subequations}
\label{eomproamp}
\begin{align}
\dot{A}_{m}(t)=&-i[\omega_{c}+m\omega_{M}+\Delta_{0}\cos(2Jt)]A_{m}(t)-iJB_{m}(t)\nonumber\\
&-ig_{0}[\sqrt{m+1}A_{m+1}(t)+\sqrt{m}A_{m-1}(t)],\\
\dot{B}_{m}(t)=&-i[\omega_{c}+m\omega_{M}-\Delta_{0}\cos(2Jt)]B_{m}(t)-iJA_{m}(t)\nonumber\\
&+ig_{0}[\sqrt{m+1}B_{m+1}(t)+\sqrt{m}B_{m-1}(t)],
\end{align}
\end{subequations}
which are solved numerically.
The fidelity between the two states $\vert\psi(t)\rangle$ and $|\psi_{\rm approx}(t)
\rangle$ can be obtained as
\begin{eqnarray}
&& F(t)=\vert\langle \psi(t)\vert \psi_{\rm approx}(t)\rangle\vert^{2}\notag \\
&& =\left\vert e^{-\frac{\vert\beta(t)\vert^{2}}{2}
}\sum_{m=0}^{\infty}\frac{\beta^{m}(t)}{\sqrt{m!}}[\cos(Jt)A_{m}^{\ast}(t)-i\sin(Jt)B_{m}^{\ast}(t)]\right\vert^{2}.\nonumber\\
\end{eqnarray}

In Fig.~\ref{fidelity}(a), we plot the fidelity as a function of
the evolution time for various values of $\Delta_{0}/g_{0}$. We
see that a good fidelity [i.e., $F(t) \approx 1$] can be obtained
for larger values of $\Delta_{0}/g_{0}$, which is in accordance
with the condition in Eq.~(\ref{conditionRWA}). The fidelity curves
exhibit some fast oscillations, which are caused by the time
factors $\exp(\pm i 2Jt)$ and $\exp(\pm i\omega_{M}t)$ ($\omega_{M}$,
$2J\gg g_{0}$). We also plot the time dependence of the mechanical
displacement $|\langle b(t)\rangle|$ in Fig.~\ref{fidelity}(b). Here
the solid curve corresponds to the analytical case $|\langle b(t)\rangle|=|\beta(t)|$,
which is independent
of $\Delta_{0}$, as given in Eq.~(\ref{detaoft}). The other three
curves in panel (b) are determined by the numerical solutions with
the corresponding values of $\Delta_{0}/g_{0}$ given in panel (a).
It can be seen that the exact numerical results match well the approximate
analytical result, except a slight shift of the period.

\begin{figure}[tbp]
\center
\includegraphics[bb=25 4 374 272, width=0.48 \textwidth]{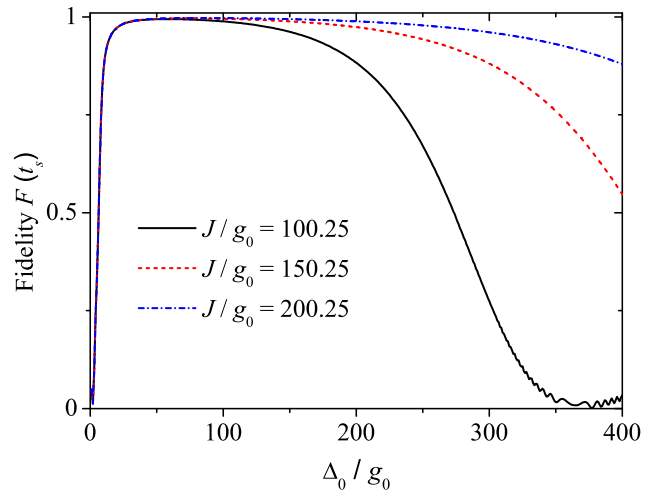}
\caption{(Color online) The fidelity $F(t_{s})$ at time $t_{s}$
versus the ratio $\Delta_{0}/g_{0}$ for various values of
$J/g_{0}$, where $t_{s}=\pi/(\omega_{M}-2J)$ is the time for the first peak
value of $|\beta(t)|$. For different values of $J$, the values of
$\omega_{M}$ are taken to keep
$\omega_{M}-2J=g_{0}/2$.}\label{fvsDelta0}
\end{figure}

Although a high ratio $\Delta_{0}/g_{0}$ would increase the
fidelity as shown in Fig.~\ref{fidelity}, there is an upper limit of $\Delta_0$
according to Eq.~(\ref{conditionRWA}). Specifically, for the approximate
Hamiltonian~(\ref{Hamiltappr1}) to be valid, we also need $\Delta_0\ll 16J/5$. In other
words, there is a finite range of $\Delta_0$ for our model to
operate with. This is illustrated in Fig.~\ref{fvsDelta0} where
the fidelity $F(t_{s})$ at $t_{s}=\pi/(\omega_M-2J)$ is plotted as
a function of $\Delta_{0}/g_{0}$. The choice of time $t_{s}$ is
useful because it is the time for the first maximum displacement.
In plotting all the curves in Fig.~\ref{fvsDelta0}, we use the
parameters with $\omega_{M}-2J=g_{0}/2$, so that the maximum
mechanical displacement is $2$. From Fig.~\ref{fvsDelta0}, we see
that fidelity is almost $1$ in a range of $\Delta_{0}/g_{0}$ when
the conditions in Eq.~(\ref{conditionRWA}) are satisfied. The
range of $\Delta_{0}/g_{0}$ giving $F(t_s) \approx 1$ becomes wider
as the ratio $J/g_{0}$ increases, as illustrated by the blue
dash-dotted curve in Fig.~\ref{fvsDelta0}.

\section{Discussions}

Our analysis so far has focused on the coherent evolution of the
modulated system. To address the effect of cavity field damping,
we let $\kappa_c$ be the cavity field damping rate for both the
left and right cavities, then with $\omega_{M}-2J \approx
g_{0}/2$, $t \approx 1/g_0$ is the time scale for the membrane
displacement being greater than the width of the mechanical ground
state. Therefore the condition $g_0 \gg \kappa_c$ would ensure
that the photon has a sufficient time to yield a significant
mechanical displacement (with $|\beta|
> 1$) before it leaks out of the cavity.

%%%%%%%%%%%%%%%%%%%%%%%%%%%%%%%%%%%%%
%%%%%%%%%%%%%%%%%%%%%%%%%%%%%%%%%%%%%

More specifically, based on the approximate Hamiltonian~(\ref{Hamiltappr2}) and
the initial state~(\ref{inistate}), it is not difficult to estimate that the
phonon number $\langle b^\dag b \rangle $ is of the order
$g_0^2/[4(\omega_M-2J)^2+ 4\kappa_c^2]$  at a time long compared
with the cavity-field lifetime, but short compared with the
lifetime of the mechanical oscillator, i.e., $ 1/\gamma_M \gg t
\gg 1/\kappa_c$, where $\gamma_M$ is the damping rate of the
membrane. Therefore, if $g_0 \gg \kappa_c$ and $\omega_M-2J <
g_0$, then the average phonon number of our modulated system can
be larger than $1$, after the photon leaks out of the cavity.

We point out that our approach can be used to create distinct
mechanical superposition
states~\cite{Armour2002,Tian2005,Liao2008,Cirac2011,Tan2013,Yin2013,Asjad2014,Ge2015}.
The procedure consists of three steps: (i) Prepare the system in
initial states $(1/\sqrt{2})(\vert 1\rangle_{L}\vert
0\rangle_{R}\pm\vert 0\rangle_{L}\vert 1\rangle_{R})\vert
0\rangle_{M}$. (ii) Then let the system evolve for a suitable time,
for example $t=2\pi/(\omega_{M}-2J)$. (iii) Measure the state of
the photon~\cite{Liao2014}. As long as the photon is measured in either the left
cavity or the right cavity, then the mechanical resonator will be
projected to a mechanical superposition state.

We note that mechanical superposed states can be
measured by using quantum state tomography~\cite{Liao2014specrec,Vanner2015}. By strongly driving another cavity mode (avoiding the cross talk with the single-photon state) at red sideband, the
optomechanical coupling can be linearized into a beam-splitter interaction. Assume a large decay of the
assistant cavity field such that it can be eliminated adiabatically, then the state of the mechanical mode can be mapped to the output light~\cite{Vitali2007}.
As a result the mechanical state can be known by reconstructing the state of the output light beam~\cite{Raymer2009}.
In addition, the technique of spectrometric reconstruction can also be used to obtain the state information of the mechanical mode by detecting the single photon spectrum~\cite{Liao2014specrec}.

Moreover, we would like to mention that the approach in this paper can
also be used to enhance the optical nonlinearities~\cite{Liaomodu}. In
optomechanics, the magnitude of the optical nonlinearity is
$g_{0}^{2}/\omega_{M}$. It can be increased by effectively
decreasing the frequency, from $\omega_{M}$ to $\omega_{M}-2J$,
under the same coupling strength $g_{0}$.

Finally, we discuss an example that illustrates the conditions
required for the experimental demonstration of the mechanism presented
in this paper. The example consists of a vibrational mode of a
silicon nitride membrane with an eigenfrequency $\omega_{M}/ 2 \pi=
1$ MHz~\cite{Regal2,Lee}, and the mechanical damping rate is less
than a few hertz and so it may be ignored. The membrane is placed in
the middle of a Fabry-Perot cavity of length $L$. A pair of nearly
degenerate normal modes of the cavity is chosen such that the
frequency splitting $2J \approx \omega_{M}$ at the avoided crossing.
To obtain mode splitting in the megahertz range, the transverse modes of
the cavity may be exploited~\cite{Sankey,Karuza2,Lee}. Consider
$L=1$ mm, then $g_0$ is about $1$ kHz, which is much smaller than
$\omega_{M}$. However, by applying a weak modulation with
$\Delta_0=40g_0 \approx 40$ kHz, the effect of radiation pressure
from a single photon can be significantly enhanced according to
our theory, provided that the cavity field damping rate $\kappa_c$
is smaller than $g_0$. Since $\kappa_c$ is of the order of few megahertz in
experiments~\cite{Regal2}, this is still a main challenge to achieve
$g_0> \kappa_c$. Note that $g_0> \kappa_c$ is also the condition for the
MIM optomechanical system to exhibit enhanced photon-photon
interactions~\cite{Ludwig2012}. Therefore the progress of designs
and fabrications of ultrahigh $Q$ cavities will be crucial to
access ultrastrong-coupling quantum effects in optomechanics.

\section{Conclusion}

In conclusion, we have proposed a method to enhance the mechanical
effects of single photons in a two-mode optomechanical system by
introducing a frequency modulation to the cavity fields.  The
quantum dynamics, which is driven by photon hopping between the
two modulated cavities and optomechanical interactions, can be
captured approximately by the Hamiltonian~(\ref{Hamiltappr2})
under the condition~(\ref{conditionRWA}). Such a Hamiltonian
indicates that a single photon can displace the membrane with
$\beta > 1$ even though $g_0/\omega_M \ll 1$. This method can be
used to create distinct superposition states of the mechanical
resonator and enhance the optical nonlinearities. Finally we
remark that these types of systems can be studied via quantum
emulations or quantum simulations~\cite{Buluta2009,Georgescu2014},
as described in~\cite{Johansson2014,Kim2015}. These describe
circuit analogs of optomechanical systems, allowing an additional
way to study the effects predicted here.

\begin{acknowledgments}
J.Q.L. would like to thank Hunan Normal University for its hospitality where part of this work was carried out.
J.Q.L. is supported by the DARPA ORCHID program through AFOSR, the National Science Foundation
under Award No. NSF-DMR-0956064, and the NSF-COINS
program under Grant No. NSF-EEC-0832819.
C.K.L. is partially supported by a grant from the
Research Grants Council of Hong Kong, Special Administrative Region of China (Project No. CUHK401812).
L.M.K. is partially supported by the National 973 Program under Grant No. 2013CB921804 and the NSF under Grants No. 11375060 and No. 11434011.
F.N. is partially supported by the
RIKEN iTHES Project, the MURI Center for Dynamic Magneto-Optics via the AFOSR Award No. FA9550-14-1-0040,
the Impact program of JST, and a Grant-in-Aid for Scientific Research (A).
\end{acknowledgments}


\begin{thebibliography}{99}
%reviews on optomechanics
\bibitem{Aspelmeyer2014}    M. Aspelmeyer, T. J. Kippenberg, and F. Marquardt, Cavity optomechanics, Rev. Mod. Phys. \textbf{86}, 1391 (2014).
\bibitem{Chen2013}          Y. Chen, Macroscopic quantum mechanics: theory and experimental concepts of optomechanics, J. Phys. B: At. Mol. Opt. Phys. \textbf{46}, 104001 (2013).

%mechanical cat state generation
\bibitem{Bose1999}   S. Bose, K. Jacobs, and P. L. Knight, Scheme to probe the decoherence of a macroscopic object, Phys. Rev. A \textbf{59}, 3204 (1999).
\bibitem{Marshall2003}  W. Marshall, C. Simon, R. Penrose, and D. Bouwmeester, Towards quantum superpositions of a mirror, Phys. Rev. Lett. \textbf{91}, 130401 (2003).

%single-photon strong coupling regime, photon blockade
\bibitem{Rabl2011}         P. Rabl, Photon blockade effect in optomechanical systems, Phys. Rev. Lett. \textbf{107}, 063601 (2011).
\bibitem{Liao2013}         J. Q. Liao and C. K. Law, Correlated two-photon scattering in cavity optomechanics, Phys. Rev. A \textbf{87}, 043809 (2013).
\bibitem{Xu2013A}          X.-W. Xu, Y.-J. Li, and Y.-X. Liu, Photon-induced tunneling in optomechanical systems, Phys. Rev. A \textbf{87}, 025803 (2013).
\bibitem{Kronwald2013}    A. Kronwald, M. Ludwig, and F. Marquardt, Full photon statistics of a light beam transmitted through an optomechanical system, Phys. Rev. A \textbf{87}, 013847 (2013).

\bibitem{Nunnenkamp2011}   A. Nunnenkamp, K. B{\o}rkje, and S. M. Girvin, Single-photon optomechanics, Phys. Rev. Lett. \textbf{107}, 063602 (2011).

\bibitem{Xu2013B}  X.-W. Xu, H. Wang, J. Zhang, and Y.-X. Liu, Engineering of nonclassical motional states in optomechanical systems, Phys. Rev. A \textbf{88}, 063819 (2013).

\bibitem{Gaofeng2013}  G. F. Xu and C. K. Law, Dark states of a moving mirror in the single-photon strong-coupling regime, Phys. Rev. A \textbf{87}, 053849 (2013).

\bibitem{Tan2014}  H. Tan,  Deterministic quantum superpositions and Fock states of mechanical oscillators via quantum interference in single-photon cavity optomechanics, Phys. Rev. A \textbf{89}, 053829 (2014).

\bibitem{Garziano2015} L. Garziano, R. Stassi, V. Macr\'{\i}, S. Savasta, and O. Di Stefano, Single-step arbitrary control of mechanical quantum states in ultrastrong optomechanics,
Phys. Rev. A \textbf{91}, 023809 (2015).

\bibitem{Liao2012}         J. Q. Liao, H. K. Cheung, and C. K. Law, Spectrum of single-photon emission and scattering in cavity optomechanics, Phys. Rev. A \textbf{85}, 025803 (2012).
\bibitem{Hu2015}            D. Hu, S. Y. Huang, J. Q. Liao, L. Tian, and H. S. Goan, Quantum coherence in ultrastrong optomechanics, Phys. Rev. A \textbf{91}, 013812 (2015).

%ground state cooling in optomechanics
\bibitem{Wilson-Rae2007}   I. Wilson-Rae, N. Nooshi, W. Zwerger, and T. J. Kippenberg, Theory of ground state cooling of a mechanical oscillator using dynamical backaction, Phys. Rev. Lett. \textbf{99}, 093901 (2007).
\bibitem{Marquardt2007}    F. Marquardt, J. P. Chen, A. A. Clerk, and S. M. Girvin, Quantum theory of cavity-assisted sideband cooling of mechanical motion, Phys. Rev. Lett. \textbf{99}, 093902 (2007).

\bibitem{Li2008}    Y. Li, Y. D. Wang, F. Xue, and C. Bruder, Quantum theory of transmission line resonator-assisted cooling of a micromechanical resonator, Phys. Rev. B \textbf{78}, 134301 (2008).

\bibitem{Tian2009}    L. Tian, Ground state cooling of a nanomechanical resonator via parametric linear coupling, Phys. Rev. B \textbf{79}, 193407 (2009).


\bibitem{Liu2015}     Y. C. Liu, R. S. Liu, C. H. Dong, Y. Li, Q. Gong, and Y. F. Xiao, Cooling mechanical resonators to the quantum ground state from room temperature, Phys. Rev. A \textbf{91}, 013824 (2015).

%increase of the ratio g0 over omega_M
\bibitem{Liaomodu}    J. Q. Liao, K. Jacobs, F. Nori, and R. W. Simmonds, Modulated electromechanics: large enhancements of nonlinearities, New J. Phys. \textbf{16}, 072001 (2014).
\bibitem{Lu2013}      X. Y. L\"{u}, W. M. Zhang, S. Ashhab, Y. Wu, and F. Nori, Quantum-criticality-induced strong Kerr nonlinearities in optomechanical systems, Sci. Rep. \textbf{3}, 2943 (2013).

\bibitem{Tang2014}   H. X. Tang and D. Vitali, Prospect of detecting single-photon-force effects in cavity optomechanics, Phys. Rev. A \textbf{89}, 063821 (2014).

\bibitem{Xuereb2012}    A. Xuereb, C. Genes, and A. Dantan, Strong coupling and long-range collective interactions in optomechanical arrays, Phys. Rev. Lett. \textbf{109}, 223601 (2012).

\bibitem{Heikkila2014}    T. T. Heikkil\"{a}, F. Massel, J. Tuorila, R. Khan, and M. A. Sillanp\"{a}\"{a}, Enhancing optomechanical coupling via the Josephson effect, Phys. Rev. Lett. \textbf{112}, 203603 (2014).

\bibitem{Rimberg2014}  A. J. Rimberg, M. P. Blencowe, A. D. Armour, and P. D. Nation, A cavity-Cooper pair transistor scheme for investigating quantum optomechanics in the ultra-strong coupling regime, New J. Phys. \textbf{16}, 055008 (2014).

\bibitem{Lu2015}  X. Y. L\"{u}, Y. Wu, J. R. Johansson, H. Jing, J. Zhang, and F. Nori, Squeezed optomechanics with phase-matched amplification and dissipation,
Phys. Rev. Lett. \textbf{114}, 093602 (2015).

\bibitem{Via2015}    G. Via, G. Kirchmair, and O. Romero-Isart, Strong single-photon coupling in superconducting quantum magnetomechanics, Phys. Rev. Lett. \textbf{114}, 143602 (2015).



%membrane-middle experiments


\bibitem{Thompson} J. D. Thompson, B. M. Zwickl, A. M. Jayich, F. Marquardt, S. M. Girvin, and J. G. E. Harris, Strong dispersive coupling of a high-finesse cavity to a micromechanical membrane, Nature (London)
{\bf 452}, 72 (2008).

\bibitem{Sankey} J. C. Sankey, C. Yang, B. M. Zwickl, A. M. Jayich, and J. G. E. Harris, Strong and tunable nonlinear optomechanical coupling in a low-loss system, Nature Phys. {\bf 6}, 707 (2010).

\bibitem{Karuza1} M. Karuza, C. Molinelli, M. Galassi, C. Biancofiore, R. Natali, P. Tombesi, G. Di Giuseppe,
and D. Vitali, Optomechanical sideband cooling of a thin membrane within a cavity, New J. Phys. {\bf 14}, 095015 (2012).

\bibitem{Regal1} T. P. Purdy, R. W. Peterson, P.-L. Yu, and C. A.
Regal, Cavity optomechanics with Si$_{3}$N$_{4}$ membranes at cryogenic
temperatures, New J. Phys. {\bf 14}, 115021 (2012).

\bibitem{Karuza2} M. Karuza, C. Biancofiore, M. Bawaj, C. Molinelli, M. Galassi, R. Natali, P. Tombesi, G. Di Giuseppe, and D.
Vitali,  Optomechanically induced transparency in a membrane-in-the-middle setup at room temperature, Phys. Rev. A {\bf 88}, 013804 (2013).

\bibitem{Karuza3} M. Karuza, M. Galassi, C. Biancofiore, C. Molinelli, R. Natali, P. Tombesi, G. Di Giuseppe, and D.
Vitali, Tunable linear and quadratic optomechanical coupling for a tilted membrane within an optical cavity: theory and experiment, J. Opt. {\bf 15}, 025704 (2013).

\bibitem{Andrews2014} R. W. Andrews, R. W. Peterson, T. P. Purdy, K. Cicak, R. W. Simmonds, C. A. Regal, and K. W. Lehnert, Bidirectional and efficient conversion between microwave and optical light, Nature Phys. \textbf{10}, 321 (2014).

\bibitem{Lee} D. Lee, M. Underwood, D. Mason, A. B. Shkarin, S. W. Hoch,  and  J. G. E. Harris, Multimode optomechanical dynamics in a cavity with avoided crossings, Nature Commun. {\bf 6}, 6232 (2015).

\bibitem{Regal2} T. P. Purdy, P.-L. Yu, N. S. Kampel, R. W. Peterson, K. Cicak, R. W. Simmonds, and C. A.
Regal, Optomechanical Raman-Ratio Thermometry, arXiv:1406.7247.

%two-mode optomechanics

\bibitem{Bhattacharya2008}   M. Bhattacharya, H. Uys, and P. Meystre, Optomechanical trapping and cooling of partially reflective mirrors, Phys. Rev. A \textbf{77}, 033819 (2008).
\bibitem{Jayich2008}    A. M. Jayich, J. C. Sankey, B. M. Zwickl, C. Yang, J. D. Thompson, S. M. Girvin, A. A. Clerk, F. Marquardt, and J. G. E. Harris, Dispersive optomechanics: a membrane inside a cavity, New J. Phys. \textbf{10}, 095008 (2008).
\bibitem{Miao2009}       H. Miao, S. Danilishin, T. Corbitt, and Y. Chen, Standard quantum limit for probing mechanical energy quantization, Phys. Rev. Lett. \textbf{103}, 100402 (2009).
\bibitem{Cheung2011}      H. K. Cheung and C. K. Law, Nonadiabatic optomechanical Hamiltonian of a moving dielectric membrane in a cavity, Phys. Rev. A \textbf{84}, 023812 (2011).
\bibitem{Ludwig2012}     M. Ludwig, A. H. Safavi-Naeini, O. Painter, and F. Marquardt, Enhanced quantum nonlinearities in a two-mode optomechanical system, Phys. Rev. Lett. \textbf{109}, 063601 (2012).
\bibitem{Stannigel2012}  K. Stannigel, P. K\'{o}m\'{a}r, S. J. M. Habraken, S. D. Bennett, M. D. Lukin, P. Zoller, and P. Rabl, Optomechanical quantum information processing with photons and phonons, Phys. Rev. Lett. \textbf{109}, 013603 (2012).
\bibitem{Liao2014}     J. Q. Liao, Q. Q. Wu, and F. Nori, Entangling two macroscopic mechanical mirrors in a two-cavity optomechanical system, Phys. Rev. A \textbf{89}, 014302 (2014).

\bibitem{Svelto}   O. Svelto, \emph{Principles of Lasers} (Plenum press, New York, 1982).


\bibitem{Schmidt2015}   M. Schmidt, S. Ke{\ss}ler, V. Peano, O. Painter, and F. Marquardt, Optomechanical creation of magnetic fields for photons on a lattice, arXiv:1502.07646.

\bibitem{Ballester2012}   D. Ballester, G. Romero, J. J. Garc\'{\i}a-Ripoll, F. Deppe, and E. Solano, Quantum simulation of the ultrastrong-coupling dynamics in circuit quantum electrodynamics, Phys. Rev. X \textbf{2}, 021007 (2012).

\bibitem{jc} When $\Delta_0=0$, the Hamiltonian~(\ref{schwingerrepH}) in the single-photon subspace can
be identified as a quantum Rabi model. Under the condition $|\omega_{M}- 2J| \le
\frac{g_0}{2}$ and $J \gg g_0$, the model can further be reduced
to the Jaynes-Cummings model by using the rotating wave approximation.
In this picture, the initial state~(\ref{inistate}) corresponds to a vacuum
field state and the two-level atom is in an equal superpositon of
excited and ground states. Therefore the atom can only deliver at
most half a quanta to the field (membrane).

%cat state generation
\bibitem{Armour2002}   A. D. Armour, M. P. Blencowe, and K. C. Schwab, Entanglement and decoherence of a micromechanical resonator via coupling to a Cooper-pair box, Phys. Rev. Lett. \textbf{88}, 148301 (2002).

\bibitem{Tian2005}   L. Tian, Entanglement from a nanomechanical resonator weakly coupled to a single Cooper-pair box, Phys. Rev. B \textbf{72}, 195411 (2005).

\bibitem{Liao2008}   J. Q. Liao and L. M. Kuang, Nanomechanical resonator coupling with a double quantum dot: quantum state engineering, Eur. Phys. J. B \textbf{63}, 79 (2008).

\bibitem{Cirac2011}    O. Romero-Isart, A. C. Pflanzer, F. Blaser, R. Kaltenbaek, N. Kiesel, M. Aspelmeyer, and J. I. Cirac, Large Quantum Superpositions and Interference of Massive Nanometer-Sized Objects, Phys. Rev. Lett. \textbf{107}, 020405 (2011).

\bibitem{Tan2013}   H. Tan, F. Bariani, G. Li, and P. Meystre, Generation of macroscopic quantum superpositions of optomechanical oscillators by dissipation, Phys. Rev. A \textbf{88}, 023817 (2013).

\bibitem{Yin2013} Z.-Q. Yin, T. Li, X. Zhang, and L. M. Duan,  Large quantum superpositions of a levitated nanodiamond through spin-optomechanical coupling, Phys. Rev. A \textbf{88}, 033614 (2013).

\bibitem{Asjad2014}    M. Asjad and D. Vitali, Reservoir engineering of a mechanical resonator: generating a macroscopic superposition state and monitoring its decoherence,  J. Phys. B: At. Mol. Opt. Phys. \textbf{47}, 045502 (2014).

\bibitem{Ge2015}     W. Ge and M. S. Zubairy, Macroscopic optomechanical superposition via periodic qubit flipping, Phys. Rev. A \textbf{91}, 013842 (2015).

%quantum state tomography
\bibitem{Liao2014specrec}  J. Q. Liao and F. Nori, Spectrometric reconstruction of mechanical-motional states in optomechanics,
Phys. Rev. A \textbf{90}, 023851 (2014).

\bibitem{Vanner2015}    M. R. Vanner, I. Pikovski, and M. S. Kim, Towards optomechanical quantum state reconstruction of mechanical motion, Ann. Phys. (Berlin) \textbf{527}, 15 (2015).

\bibitem{Vitali2007}   D. Vitali, S. Gigan, A. Ferreira, H. R. B\"{o}hm, P. Tombesi, A. Guerreiro, V. Vedral, A. Zeilinger, and M. Aspelmeyer, Optomechanical entanglement between a movable mirror and a cavity field, Phys. Rev. Lett. \textbf{98}, 030405 (2007).

\bibitem{Raymer2009}   A. I. Lvovsky and M. G. Raymer, Continuous-variable optical quantum-state tomography, Rev. Mod. Phys. \textbf{81}, 299 (2009).


%F Nori's papers
\bibitem{Buluta2009}  I. Buluta and F. Nori, Quantum simulators, Science \textbf{326}, 108 (2009).
\bibitem{Georgescu2014}   I. M. Georgescu, S. Ashhab, and F. Nori, Quantum simulation, Rev. Mod. Phys. \textbf{86}, 153 (2014).

\bibitem{Johansson2014}  J. R. Johansson, G. Johansson, and F. Nori, Optomechanical-like coupling between superconducting resonators, Phys. Rev. A \textbf{90}, 053833 (2014).
\bibitem{Kim2015}   E.-J. Kim, J. R. Johansson, and F. Nori, Circuit analog of quadratic optomechanics, Phys. Rev. A \textbf{91}, 033835 (2015).

\end{thebibliography}
\end{document}